\documentclass[journal]{IEEEtran}

\usepackage{setspace}
\usepackage{amsmath}
\usepackage{graphicx}
\usepackage[linesnumbered,ruled,vlined]{algorithm2e}
\usepackage{multirow}
\usepackage{cite}

\usepackage{lipsum}
\usepackage{fancyhdr}
\begin{document}
\title{Swarm Intelligence Based Multi-phase OPF For Peak Power Loss Reduction In A Smart Grid}
\author{\authorblockN{Adnan Anwar, A. N. Mahmood}
\thanks{Adnan Anwar, A. N. Mahmood are with the University of New South Wales, Canberra. (E-mail: Adnan.Anwar@adfa.edu.au; A.Mahmood@adfa.edu.au)}}
\maketitle

\begin{abstract}


Recently there has been increasing interest in improving smart grids efficiency using computational intelligence. A key challenge in future smart grid is designing Optimal Power Flow tool to solve important planning problems including optimal DG capacities.
Although, a number of OPF tools exists for balanced networks there is a lack of research for unbalanced multi-phase distribution networks. In this paper, a new OPF technique has been proposed for the DG capacity planning of a smart grid. During the formulation of the proposed algorithm, multi-phase power distribution system is considered which has unbalanced loadings, voltage control and reactive power compensation devices. The proposed algorithm is built upon a co-simulation framework that optimizes the objective by adapting a constriction factor Particle Swarm optimization. The proposed multi-phase OPF technique is validated using IEEE 8500-node benchmark distribution system.

\end{abstract}
\IEEEpeerreviewmaketitle

\section{Introduction}


Traditionally, a utility power distribution systems is designed as a passive network which only allows power to flow from upstream to downstream where it is assumed that a primary substation is the
sole source of power. To mitigate the problem of ever growing load demand and to increase the distribution network efficiency, significant amount of Distributed Generation (DG) units are being integrated in the low voltage distribution system which would make the system active by introducing bi-directional power flows~\cite{Keane1}. In a smart grid environment, this active distribution system with distributed energy resources (DERs) needs advanced functionalities and analysis tools for planning as well as real-time operations~\cite{DuganSimcity}. In a smart grid environment, efficient and reliable power delivery to the end-users is a major concern. To enhance the efficiency,
accurate modeling of distribution system including the effects of distributed energy resources (DERs) is important. A co-simulation framework can be useful for advanced modeling of the future smart grid applications~\cite{Taylor6038952,Godfrey5622057}.


%
%

Optimal Power Flow (OPF) is a powerful tool which is widely used for different extended real-time operation and medium-long term planning~\cite{Bruno1}.
Traditionally, OPF is a static non-linear analysis tool which is used for solving different power system optimization problems under network constraints. Different methods are widely
used for solving OPF problems which includes mathematical techniques, heuristics and meta-heuristics. Examples of mathematical OPF techniques used in power system are
linear programming, quadratic programming, non-linear programming, the interior point method~\cite{Li6204238,Momoh336133}. Different meta-heuristic techniques are also widely employed for solving OPF problems. These types of meta-heuristic techniques do not need to determine the gradient or Hessian matrix of the function which needs to be optimized; therefore,
can be adopted for those cases where the optimization problem is not differentiable and irregular.
Moreover, mathematical optimization techniques (i.g., Newton's method or Quasi-Newton's method) which make the uses of gradients, finds the stationary point of a function where the value of the gradient is zero or very close to zero. One limitation of gradient based methods is that they linearize the objective function and the system constraints around an operating point which make a high probability of converging quickly to one of the local minima~\cite{Kolda03optimizationby,Abido2002563}. On the contrary, evolutionary computation based methods are much more effective for achieving solutions on a rough or complex solutions surface by avoiding local minima~\cite{park2005,Abido2002563}.
%

Most of the existing OPF techniques found in the literature consider a balanced approximation of the power system~\cite{Niknam6213709,Kumari2010736,Sojoudi6039236}, i.e., in those cases, it is assumed that the transmission lines are transposed and the three-phase loading is balanced. However, in a practical distribution system, power delivery lines are not transposed and loads are not balanced~\cite{Kerstingrecomm}. Often, these unbalanced distribution systems are also multi-phase in nature, i.e, power delivery lines have the mix of single-phase, two-phase, three-phase and/or neutral lines. Hence, OPF considering multi-phase unbalanced test system, which we define as multi-phase OPF, offers more accurate solutions for the distribution system analysis. Besides, widespread use of single-phase
distributed generation (DG) units like solar PV cell and small wind turbines have made the grid more prone to voltage imbalance~\cite{Bruno1}. Therefore, an accurate OPF technique should consider unbalanced and multi-phase nature of the distribution system.
Although significant amount of research has been done for developing
efficient and fast techniques of OPF~\cite{Niknam6213709,Kumari2010736,Sojoudi6039236} for traditional balanced networks, there is a lack of studies on OPF techniques for multi-phase distribution system analysis of the future smart grids~\cite{Bruno1}.
In this paper, IEEE benchmark 8500 node multi-phase distribution test system is used and Electric Power Research Institute's smart grid tool OpenDSS is used to model multi-phase unbalanced test system. A case study based on Multi-phase OPF is performed to determine optimal DG capacities for loss reduction. To achieve a better converge characteristic in a complex environment during DG capacity allocation, Constriction Factor PSO (CF-PSO) is used for optimization purpose.

The organization of this paper is as follows- In Section~\ref{psos}, concepts and description of basic PSO and CF-PSO is discussed.
The problem statement of this paper is given in Section~\ref{PStatement}. The problem is formulated in Section~\ref{Pform}. The architecture of the solution methodology is given in Section~\ref{SFnTS} where the description of test system is also discussed.
The algorithm is discussed in Section~\ref{SecSA} and the experimental results and analysis are presented in Section~\ref{SecRes}.
Finally, the paper concludes with some brief remarks in Section~\ref{SecEnd}.

\section{PSO Formulation: Parameters and Variants}\label{psos}


The algorithm is initialized by generating random population which is referred as a swarm. The dimension of the swarm depends on the problem size.
In a swarm, each individual possible solution is represented as a `particle'. At each iteration, positions and velocities of particles are updated depending on their individual and collective behavior.
At the first step of the optimization process, an \emph{n}-dimensional initial population (swarm) and control parameters are initialized. Each particle of a swarm is
associated with the position vector, $\textbf{x}_{i} = [{x}_{i}^{1},{x}_{i}^{2},...,{x}_{i}^{n}]$  and the velocity vector, $\textbf{v}_{i} = [{v}_{i}^{1},{v}_{i}^{2},...,{v}_{i}^{n}]$,
where n represents the search space dimension. Before going to the basic PSO loop, the position and velocity of each particle is initialized. Generally, the initial position of the ${i}^{th}$ particle
${x}_{i}$ can be obtained from uniformly distributed random vector U (${x}_{min}, {x}_{max}$), where ${x}_{min}$ and ${x}_{max}$ represents the lower and upper limits of the solution space respectively.
During the optimization procedure, position of each particle is updated using~(\ref{peq})
\begin{equation}
\textbf{x}_{i}^{t+1} = \textbf{x}_{i}^t+\textbf{v}_{i}^{t+1}
\label{peq}
\end{equation}
where ${x}_{i}\in{R}^{n}$ and ${v}_{i}\in{R}^{n}$.
\\
At each iteration, new velocity for each particle is updated which drives the optimization process. The new velocity of any particle is calculated based on
its previous velocity, the particle's best known position and the swarm's best known position. Particle's best known position is it's
location at which the best fitness value so far has been achieved by itself and swarm's best known position is the location at which the best fitness value
so far has been achieved by any particle of the entire swarm~\cite{EberhartConsPSO}. The velocity equation drives the optimization process which is updated using~(\ref{veq})
\begin{equation}                                                                                                                                                                                    \textbf{v}_{i}^{t+1} = w.\textbf{v}_{i}^t+ {r}_{1}.{c}_{1}.(\textbf{p}_{i}-\textbf{x}_{i}^t)+ {r}_{2}.{c}_{2}.(\textbf{p}_{g}-\textbf{x}_{i}^t)
\label{veq}
\end{equation}
In this equation, \emph{w} is the inertia weight. $(\textbf{p}_{i}-\textbf{x}_{i}^t)$ represents the
`self influence' of each particle which quantifies the performance of each particle with it's previous performances. The component $(\textbf{p}_{g}-\textbf{x}_{i}^t)$ represents
the `social cognition' among different particles within a swarm and quantify the performance relative to other neighboring particles.
The learning co-efficients ${c}_{1}$ and ${c}_{2}$ represent the trade-off between the self influence part and the social cognition part of the particles~\cite{Zeineldin2011}.
The values of ${c}_{1}$ and ${c}_{2}$ are adopted from previous research and is typically set to 2~\cite{Etemadi1}.
In eqn~(\ref{veq}), ${P}_{i}$ is particle's best known position and ${P}_{g}$ is swarm's best known position.

In the solution loop of PSO, the algorithm continues to run iteratively, until one of the stopping conditions is satisfied~\cite{engelbrecht2007computational}.

\subsection{Constriction factor PSO (CF-PSO) with boundary conditions}

To achieve better stability and convergent behavior of PSO, a constriction factor has been introduced by Clerc and Kennedy in~\cite{Clerc2002}. The superiority of CF-PSO over inertia-weight
PSO is discussed in~\cite{EberhartConsPSO}. Basically, the search procedure of CF-PSO is improved using the eigenvalue analysis and the system behavior can be controlled which ensures a
convergent and efficient search procedure~\cite{Vlachogiannis2006}. To formulate CF-PSO, (\ref{veq}) is replaced by~(\ref{veqcons})-(\ref{kcons2})\cite{Valle1,Naka2003}.
\begin{equation}
\textbf{v}_{i}(t) = \emph{k}~[\textbf{v}_{i}^t+ {r}_{1}.{c}_{1}.(\textbf{p}_{i}-\textbf{x}_{i}^t)+ {r}_{2}.{c}_{2}.(\textbf{p}_{g}-\textbf{x}_{i}^t)]
\label{veqcons}
\end{equation}
where
\begin{equation}
\emph{k}= \frac{2}{| ~2- \varphi - \sqrt {\varphi^2-4\varphi} ~|}
\label{kcons}
\end{equation}
and
\begin{equation}
\varphi=  {c}_{1} +{c}_{2},~ ~   \varphi > 4
\label{kcons2}
\end{equation}
here the value of $\varphi$ must be greater than 4 to ensure a stable and convergent behavior~\cite{EberhartConsPSO,Clerc2002}. Usually, the value of $\varphi$ is set to 4.1 (${c}_{1} ={c}_{2}=2.05$); therefore, the value of \emph{k} becomes 0.7298~\cite{Naka2003}.

%

\section{Problem Statement}\label{PStatement}

The objective of this analysis is to determine optimal capacities of DG units. To reduce a utility's operating costs and enhance its efficiency, it is important to reduce power losses in the distribution area. To unload a line and to reduce utility's operating costs, the integration of DG into the distribution level is a good solution. Moreover, reduction of power loss may have some positive impact on the feeder by reducing the voltage drop and improving the voltage profile of
the system~\cite{Atwa2}. However, the reverse power flow due to an excessive DG capacity may increase total circuit
losses~\cite{Anwar1}. Besides, power delivery elements can be overheated due to increased power flow the through the network, which can also decrease their reliabilities. The main objective of this work is to show how the proposed UM-OPF technique can use the mixture of both three-phase and two phase DG units to determine the minimum active power loss profile, as well as identifying the optimal DG capacities of the network. The objective function for this test case can be expressed as
\begin{equation}\label{Losstotallbl}
  min ~~~ C_{obj2} = (P_{Loss})^2
\end{equation}
where
\begin{equation}
{P_{Loss}}=\sum_{i,j=1;i\neq j}^n P_{ij}=\sum_{i,j=1;i\neq j}^n \Re~(V_i I_{ij}^* )~;~~~i~\forall~N
\end{equation}
here $P_{ij}$ is the power loss in a power delivery line between node $i$ and $j$ and `\emph{N}' is the total number of nodes.
According to Kersting in~\cite{kersting2002distribution}, a straightforward application of the $I^2R$ loss formula is not valid for multi-phase distribution system loss calculation. Hence the real power loss of a line segment can be calculated using the summation of the incoming and outgoing power in that line section. The total power loss of the system includes the total power loss of the line segments and the transformer losses~\cite{kersting2002distribution}.

In this analysis the static voltage stability margin is also considered by introducing the constraint below:
\begin{equation}
  V_{min} \leq V_{i} \leq V_{max}~; ~~~i~\forall~N    \label{vequchp}
\end{equation}
where $V_{min}$ and $V_{max}$ are the minimum and maximum voltage limits respectively. To maintain a stable operation, voltage limit needs to be within $\pm 6\%$ of the nominal voltage~\cite{Masters1}.

Current passing through each of the power delivery lines must not exceed it's maximum rating. The value of maximum current carrying capacity ($I_{ij}^{max}$) of any conductor, known as ampacity, can be found in~\cite{kersting2002distribution}.
\begin{equation}\label{ilimit}
I_{ij} < I_{ij}^{max}~;  ~~~~~~i,j~\forall~N
\end{equation}

Now, the search space is limited by the DG capacity limits bounded by Eqn. (\ref{DGlimit}). Here ${P}_{min}^{DG}$ and   ${P}_{max}^{DG}$ denotes the minimum and maximum expected DG capacity in kW respectively, and ${P}_{i}^{DG}$ is the optimal DG capacity which needs to be determined.
\begin{equation}
  {P}_{min}^{DG} \leq {P}_{i}^{DG} \leq {P}_{max}^{DG}~; ~~~i~\forall~{DG~units} \label{DGlimit}
\end{equation}

\begin{figure}
    \centering
    \includegraphics[width=3.4in,height=3.6in]{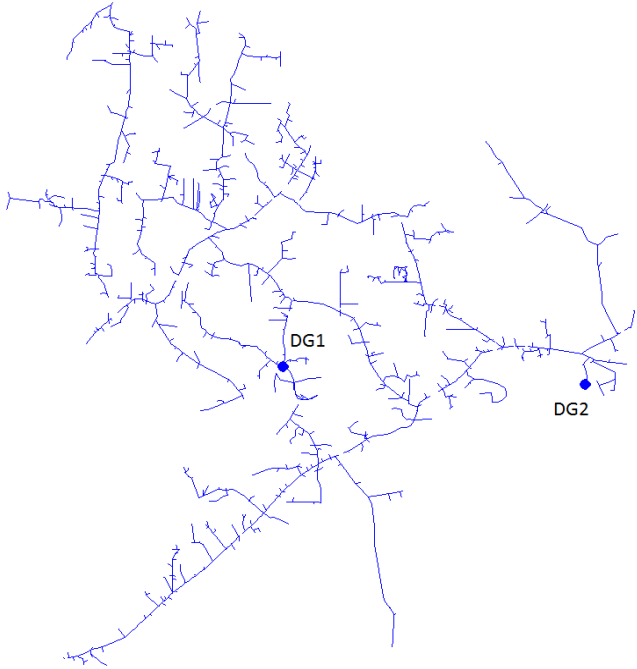}
    \caption{8500 node test feeder with DG}
    \label{fig:8500nodewithDGlbl}
\end{figure}

\section{Problem Formulation}\label{Pform}
The formulation of UM-OPF is quite similar to the
single-phase OPF except the formulation of the steady-state distribution network equations~\cite{Bruno1}. Therefore, the
minimization problem of UM-OPF can be formulated as~\cite{Bruno1}:
\begin{equation}
{\rm{min}}~~{f}_{obj}(\textbf{x},\textbf{u})
\end{equation}
subject to
\begin{equation}
\textbf{g}(\textbf{x},\textbf{u})=0\label{eqcons}
\end{equation}
\begin{equation}
\textbf{h}(\textbf{x},\textbf{u})\leq0 \label{ineqcons}
\end{equation}
where
\begin{equation}
\textbf{x}\in{\Re}^{n}~\rm{and}~\textbf{u}\in{\Re}^{m}\nonumber
\end{equation}
Here ${f}_{obj}$ is the objective function which needs to be minimised, $\textbf{x}$ is the vector of dependent variables, $\textbf{u}$ is the vector of independent variables which are generally different power system parameters. In~(\ref{eqcons}), $\textbf{g}$ is the set of equality constraints which are basically load-flow equations and $\textbf{h}$ represents the set of inequality constraints shown in~(\ref{ineqcons}). Here, the feasible solution domain of the input variables can be controlled by the following constraints:
\begin{equation}\label{ineqinput}
\textbf{u}_{min} \leq \textbf{u} \leq \textbf{u}_{max}
\end{equation}
Here, $\textbf{u}$ represents active power injection by the DG sources. In this test case,

one three-phase synchronous DG is installed at the bus \textit{m1069376} and another two-phase DG unit is installed in the low voltage side at bus \textit{SX3234149B} as shown in Fig.~\ref{fig:8500nodewithDGlbl}.
The objective of this test case is to determine the optimal generation capacities of these two DG units, therefore, $\textbf{u} = [P_{DG1}^{3-\phi}~,~P_{DG2}^{2-\phi}]$. The optimal capacities of the DG units are limited following~(\ref{DGlimit}) using the boundary conditions of the CF-PSO. In~(\ref{DGlimit}), the value of ${P}_{min}^{DG}$ and ${P}_{max}^{DG}$ depends on the planner's requirements. Here, it has been considered that the three-phase optimal DG can be any rating within the range of 10 kW to 3000 kW and two-phase DG is between zero to 500 kW. All two DG sources are synchronous type DG which are connected at 12.47 kV ans 0.208 kV level at unity pf respectively.

\section{Solution Framework and Test System} \label{SFnTS}
\begin{figure}
    \centering
    \includegraphics[width=3.4in,height=2.4in]{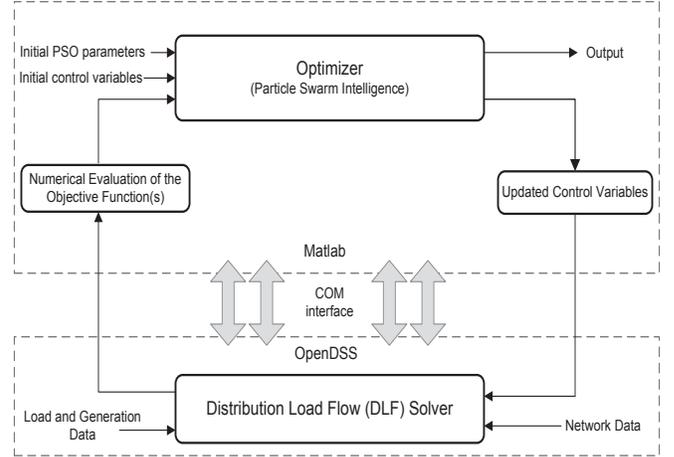}
    \caption{Co-Simulation platform for UM-OPF}
    \label{solstructure}
\end{figure}
The proposed analysis schema of UM-OPF uses a co-simulation framework where CF-PSO is used for optimization purpose and OpenDSS is used for the distribution load-flow (DLF).
The solution structure is shown in Fig.~\ref{solstructure} where the CF-PSO and the DLF have been implemented separately although they interact with each other at every iteration. The solution algorithm
of CF-PSO is implemented using Matlab and OpenDSS is interfaced with Matlab using in-process Component Object Model (COM) server Dynamic-link library (DLL). One advantage of this type of co-simulation platform is that, both the power distribution system and the optimization algorithm can be modeled in a detailed manner.
IEEE has developed several test distribution systems with unbalanced loadings, mixture of different phases (single,~two~or multi-phase power delivery lines), reactive power compensation devices, voltage and current control devices. In this research work, IEEE benchmark 8500 node test system~\cite{Arritt5484381} is used.
This test feeder has 170km of primary (MV) conductor where the maximum distance of the load node from the substation is
approximately 17km. The circuit has four capacitor banks which are turned on during the peak hours. To maintain the voltage profile, one voltage regulator is put at the substation and four other along the line~\cite{Arritt5484381}.
 Detail data of the test system can be obtained from~\cite{radialtf}.
\section{Solution Algorithm} \label{SecSA}

The step-by-step procedure of the solution is described briefly in Algorithm 1
\IncMargin{.4em}
\begin{algorithm}[!h]
\SetAlgoLined
\SetKwInOut{Input}{input}
\Input{Objective function $C_{tobj}$, swarm size $M$, CF-PSO parameters}

\For{each particle $i=1,...,M$ }{
Initialize particle's position $\textbf{x}_{i}$\;
Initialize particle's velocity $\textbf{v}_{i}$\;
}

Solve DLF to calculate $C_{tobj}$\;

Initialize particle's best known position $\textbf{P}_{i}$\;
Initialize swarm's best known position $\textbf{P}_{g}$\;

\While{stopping criterion is false}{

\For{each particle $i=1,...,M$ }{

Update particle's velocity $\textbf{v}_{i}$\;
Update particle's position $\textbf{x}_{i}$\;

Solve DLF to calculate $C_{tobj}$\;

\If{$C_{tobj}(\textbf{x}_{i}) < C_{tobj}(\textbf{P}_{i})$ }{
   Update particle's best known position $\textbf{P}_{i}$ \;
    \If{$C_{tobj}(\textbf{P}_{i}) < C_{tobj}(\textbf{P}_{g})$ }{
      Update swarm's best known position $\textbf{P}_{g}$\;
   }
}
}
}
\caption{}
\end{algorithm}\DecMargin{0.4em}

\section{Results and Discussion}\label{SecRes}

\begin{figure}
    \centering
    \includegraphics[width=3.4in]{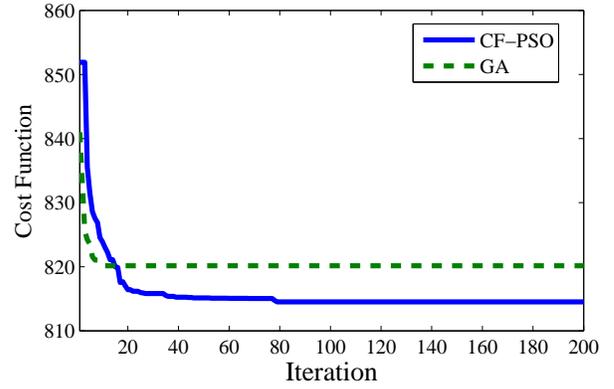}
    \caption{Convergence behavior of the CF-PSO and GA based UM-OPF}
    \label{convchar}
\end{figure}


\begin{figure}
    \centering
    \includegraphics[width=3.5in]{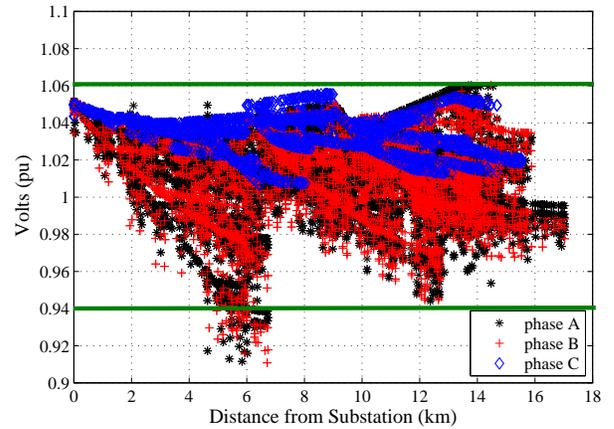}
    \caption{Voltage profile of 8500 node test system at base case}
    \label{vprof}
\end{figure}

\begin{figure}
    \centering
    \includegraphics[width=3.5in]{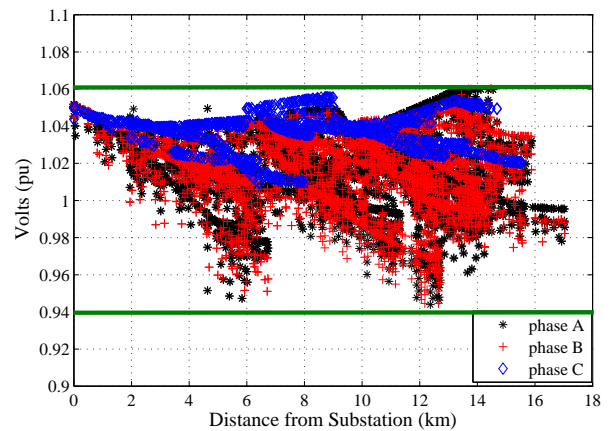}
    \caption{Voltage profile of 8500 node test system using optimal DG}
    \label{vprofDG}
\end{figure}


The optimal DG capacities of the two DG units obtained using the proposed method, are shown in Table~\ref{resultDG}. The capacity of the 3-phase DG unit is 2633 kW, and the capacity of the two-phase DG unit is 157 kW. To check the validity of the proposed method, Genetic Algorithm (GA) based multi-phase OPF is also used. Results obtained from the GA based method is also presented in Table~\ref{resultDG}. Both of the method produces almost same results, however, CF-PSO based multi-phase OPF finds better accuracy which is discussed below.

For the base case, total distribution system power demand by the load is 10.773 MW and active power loss is 1.272 MW which is around 11.8 \% of the load demand. Installing the DG units with optimal capacities as shown in Table~\ref{resultDG}, loss reduces to 0.814 MW  which is only 7.56\% as given in Table~\ref{impactDG}. Therefore, base case loss is reduced around 3.8\%. With the default configurations, voltage drop problem occurs as shown in Fig.~\ref{vprof} although voltage control devices are used. After installing the DG with optimal capacities, voltage profile is improved noticeably and no node voltage violates the stability limit as given in Fig.~\ref{vprofDG}.

\begin{table}

\caption{Optimal Generation Capacity of Two DG Units}\label{resultDG}
\centering
\small
\begin{center}
\begin{tabular}{|c|c|c|c|}
  \hline
Unbalanced  & Capacity of  & Capacity of  \\
Multi-phase & DG1 (MW) & DG2 (MW)  \\
OPF& @ m1069376 & @ SX3234149B \\

 \hline
Co-Sim (CF-PSO) & 2.633 & 0.157 \\ \hline
Co-Sim (GA) & 2.625 & 0.082  \\ \hline

\end{tabular}
\end{center}
\end{table}

\begin{table}

\caption{Comparison of the optimal employment of DG with base case }\label{impactDG}
\centering
\footnotesize
\begin{center}
\begin{tabular}{|c|c|c|c|}
  \hline
Scenario &  Base & Multi-phase &Multi-phase    \\
 &   Case &OPF (GA) &OPF (CF-PSO)   \\  \hline
Line Loss (MW) &1.086&0.665&\textbf{0.659}\\  \hline
Transformer Loss (MW)&0.185&0.155&\textbf{0.154}\\  \hline
Total Loss (MW)&1.272&0.820&\textbf{0.814}\\  \hline
Load Power (MW)&10.773&10.773&\textbf{10.773}\\  \hline
\%~Loss&11.81 \%&7.62\%&\textbf{7.56\%}\\  \hline


\end{tabular}
\end{center}
\end{table}

\section{Conclusion}\label{SecEnd}


The proposed UM-OPF technique, based on the co-simulation framework is very flexible and efficient. In the current solution framework, detailed modeling is possible for power system DLF and optimization as they are implemented separately. In this paper, optimal capacities of DG units are determined based on the proposed framework. By installing DG units with optimal capacities, the total distribution loss is reduced by 452 kW which is around 3.7\% reduction of the total system loss. The results obtained from the proposed method is also verified using a GA based method. It is desirable that the proposed methodology will have a significant impact on the planning of a smart grid.



{}


\end{document}